\documentstyle[11pt,rafringe,twoside,epsfig,makeidx]{article}
\makeindex

\def\edcomment#1{\iffalse\marginpar{\raggedright\sl#1\/}\else\relax\fi}
\reversemarginpar

\begin{document}
\title{Ionised, Atomic, and Molecular Gas around the Twin Radio Jets of NGC\,1052}
\author{R. C. Vermeulen}
\affil{Astron, P.O. Box 2, NL--7990 AA Dwingeloo, The Netherlands}
\author{E. Ros, J. A. Zensus}
\affil{Max-Planck-Institut f\"ur Radioastronomie, Auf dem H\"ugel 69,
       D-53121 Bonn, Germany}
\author{K. I. Kellermann}
\affil{National Radio Astronomy Observatory, 520 Edgemont Road,
       Charlottesville, VA 22903, USA}
\author{M. H. Cohen}
\affil{California Institute of Technology, Pasadena, CA 91125, USA}
%\author{J.A. Zensus}
%\affil{Max-Planck-Institut f\"ur Radioastronomie,
%       Auf dem H\"ugel 69, D-53121 Bonn, Germany}
\author{H. J. van~Langevelde}
\affil{Joint Institute for VLBI in Europe, P.O. Box 2, NL--7990 AA
       Dwingeloo, The Netherlands}

\begin{abstract}
  The bright radio structure of the LINER elliptical galaxy \Index{NGC\,1052}
  is dominated by bi-symmetric jets on parsec scales. Features move
  outward on both sides of the core at $v_{\rm app}\sim0.26c$. 
We have established the occurrence of free-free absorption, and suggest
the presence of a patchy, geometrically
%  There is free-free absorption, probably from a patchy, geometrically 
thick
  region oriented roughly orthogonal to the jets; components have a
  wide range of spectral shapes and brightness temperatures. We
  distinguish three velocity systems in H\,{\sc i} absorption. The ``high
  velocity'' system is the most prominent of these; it is redshifted
  from systemic by about 150\,km\,s$^{-1}$. In H\,{\sc i} VLBI it is seen
  towards both jets, but appears to be restricted to a shell 1 to 2\,pc
  away from the core. The central hole might be largely ionised, and
  could be connected to the free-free absorption. WSRT spectroscopy
  shows 1667 and 1665\,MHz OH main line absorption over at least the
  full $\sim$250\,km\,s$^{-1}$ velocity range seen in H\,{\sc i}.  In the
  ``high velocity'' system, the profiles of the OH main lines and H\,{\sc
  i} are similar, which suggests co-location of molecular and atomic
  gas. The OH satellite lines are also detected in the ``high velocity''
  system: 1612\,MHz in absorption and 1720\,MHz in emission, with
  complementary strength. But we have no satisfactory model to explain
  all properties; the connection to H$_2$O masing gas at the same
  velocity but apparently a different location is also unclear.
\end{abstract}

%\section{Introduction}

\Index{NGC\,1052}, an elliptical LINER galaxy (e.g., Gabel et al.\ 2000), is at
a distance of only 22\,Mpc (assuming ${\rm H}_\circ$ =
65\,km\,s$^{-1}$\,Mpc$^{-1}$ and using $cz=1474$\,km\,s$^{-1}$ from
Sargent et al.\ 1977), so that detailed sub-parsec scale scrutiny is
possible with VLBI (1\,mas $\approx$ 0.1\,pc). The unusually bright (1
to 2\,Jy) radio structure is core-dominated, and has two lobes with a
span of only about 3\,kpc (Wrobel 1984).
The overall spectrum is fairly flat and has sometimes been
classified as Gigahertz peaked (e.g., de Vries, Barthel, \& O'Dea
1997).

\begin{figure}[tb!]
%\plotfiddle{vermeulen_ngc1052_fig1.ps}{6cm}{0}{67}{67}{-218}{-200}
\begin{center}
\vspace{-10pt}
\includegraphics[clip,width=0.9\textwidth]{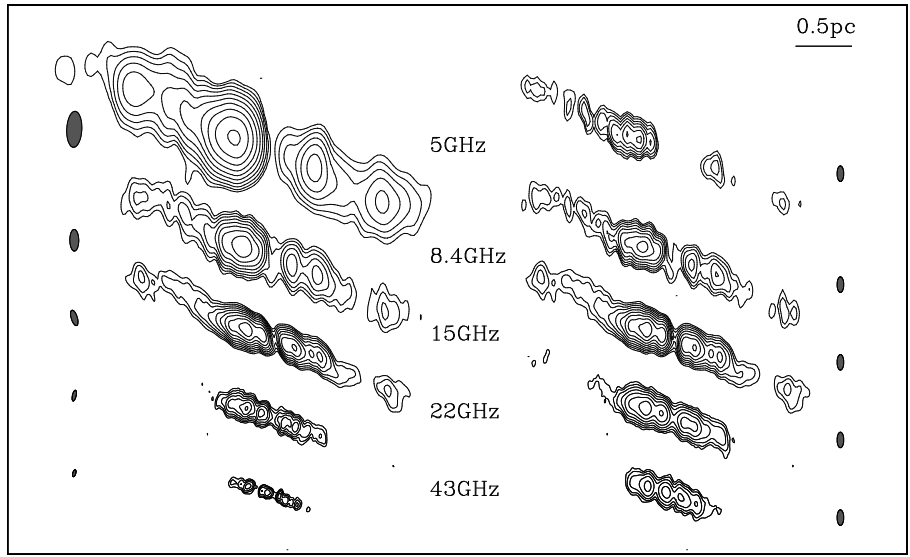}
\vspace{-15pt}
\end{center}
\caption{Contour images of \Index{NGC\,1052} at 43, 22, 15, 8, and 5\,GHz,
observed at epoch 1997.52. The images on the left-hand side are at a
resolution appropriate to the frequency (see restoring beam shown). To
facilitate inter-comparison, the images in the right-hand column have a
fixed restoring beam of 0.14$\times$0.06\,pc. The lowest contour levels
are 0.83, 1.21, 1.57, 3.44, and 1.04\,mJy\,beam$^{-1}$, and in all cases
the contours increase by factors of $2^{1/2}$.
\textsl{(Taken from Vermeulen et al.\ 2003, A\&A, 401, 113.)}
}
\end{figure}

Fortunately, given all of the valuable information which has now emerged
about it, \Index{NGC\,1052} was bright enough to be included in the 15\,GHz VLBA
survey (Kellermann et al.\ 1998), which was initiated by Ken when it
became possible to undertake such a substantial effort at a relatively
high frequency. The first 15\,GHz VLBA observation, in July 1995,
%revealed that, \Index{NGC\,1052} has a bi-symmetric morphology (see also
%Fig.~1). This is of course rather unusual. We therefore not only
%included 
revealed that, in contrast to most of the sources in the sample,
\Index{NGC\,1052} has a bi-symmetric morphology (see also Fig.~1). We therefore
not only included
\Index{NGC\,1052} in many of our regular VLBA 15\,GHz monitoring
observations, but also initiated dedicated observations of this
source. The first report on those was given by Ken, and included our
discovery of two-sided motions as well as of free-free absorption
(Kellermann et al.\ 1999). We have since then written a comprehensive
paper on all of our findings (Vermeulen et al.\ 2003). The main results
can be summarised as follows:

\paragraph{1.} In the slightly curved bi-symmetric jets (e.g., Fig.~1), multiple
sub-parsec scale features have been traced in ten 15\,GHz VLBA epochs
over five years. They move outward, reasonably linearly, at typically
$v_{\rm app}\sim0.26c$ on both sides. Given their symmetry, the jets
must be oriented near the plane of the sky.

\paragraph{2.} Nearly simultaneous VLBA observations were made at up to seven
frequencies; some of the images are displayed in Figure~1. They show a
central gap in the radio emission, which widens at lower radio
frequencies. Detailed spectra demonstrate that it is due to free-free
absorption. The western jet is covered more extensively, showing
partial obscuration along at least 1\,pc; this must be the receding
jet. But the eastern jet is also partially covered, along approximately
0.3\,pc, and so the free-free absorber is probably geometrically
thick, and oriented roughly orthogonal to the jets. The ionised gas
near the core, if distributed uniformly along a path-length of 0.5\,pc,
would have a volume density of $n_{\rm e}\sim10^5$\,cm$^{-3}$. The
opacity globally decreases away from the centre, but the variability of
moving components, acting as ``probes behind the screen'', suggests
considerable patchiness, and the wide range of component spectral
shapes suggests that synchrotron self-absorption and free-free
absorption both play a role in the inner parsecs of the jets.

\paragraph{3.} VLBI observations show that H\,{\sc i} absorbing atomic gas is
distributed in front of the approaching as well as the receding jet,
and has sub-pc scale structure. This is illustrated in Figure~2. We
distinguish three absorption systems with different characteristics.
From their substructure, at least two of them are situated close to the
AGN rather than in the galaxy as a whole. The most prominent absorption
system, with a peak optical depth of 20--25\%, is at ``high velocity'',
receding by 125--200\,km\,s$^{-1}$ with respect to the systemic
velocity. Under the uncertain assumptions that this atomic gas has a
spin temperature $T_{\rm sp}=100$\,K, and uniformly fills a
path-length of 0.5\,pc, we derive a column depth of $N_{\rm
H}\sim10^{21}$\,cm$^{-2}$, and a density of $n_{\rm H} \sim
1000$\,cm$^{-3}$. This absorber may have a continuous velocity gradient
of some 10\,km\,s$^{-1}$\,pc$^{-1}$ across the nucleus, but while it is
seen at a distance of 1--1.5\,pc on both sides of the centre, there is
a deficit in the innermost parsec; this ``central hole" in atomic gas
may be largely ionised.

\begin{figure}[htb!]
%\plotfiddle{vermeulen_ngc1052_fig2.ps}{7cm}{0}{40}{30}{-130}{-10}
\begin{center}
\vspace{-10pt}
\includegraphics[clip,width=0.80\textwidth]{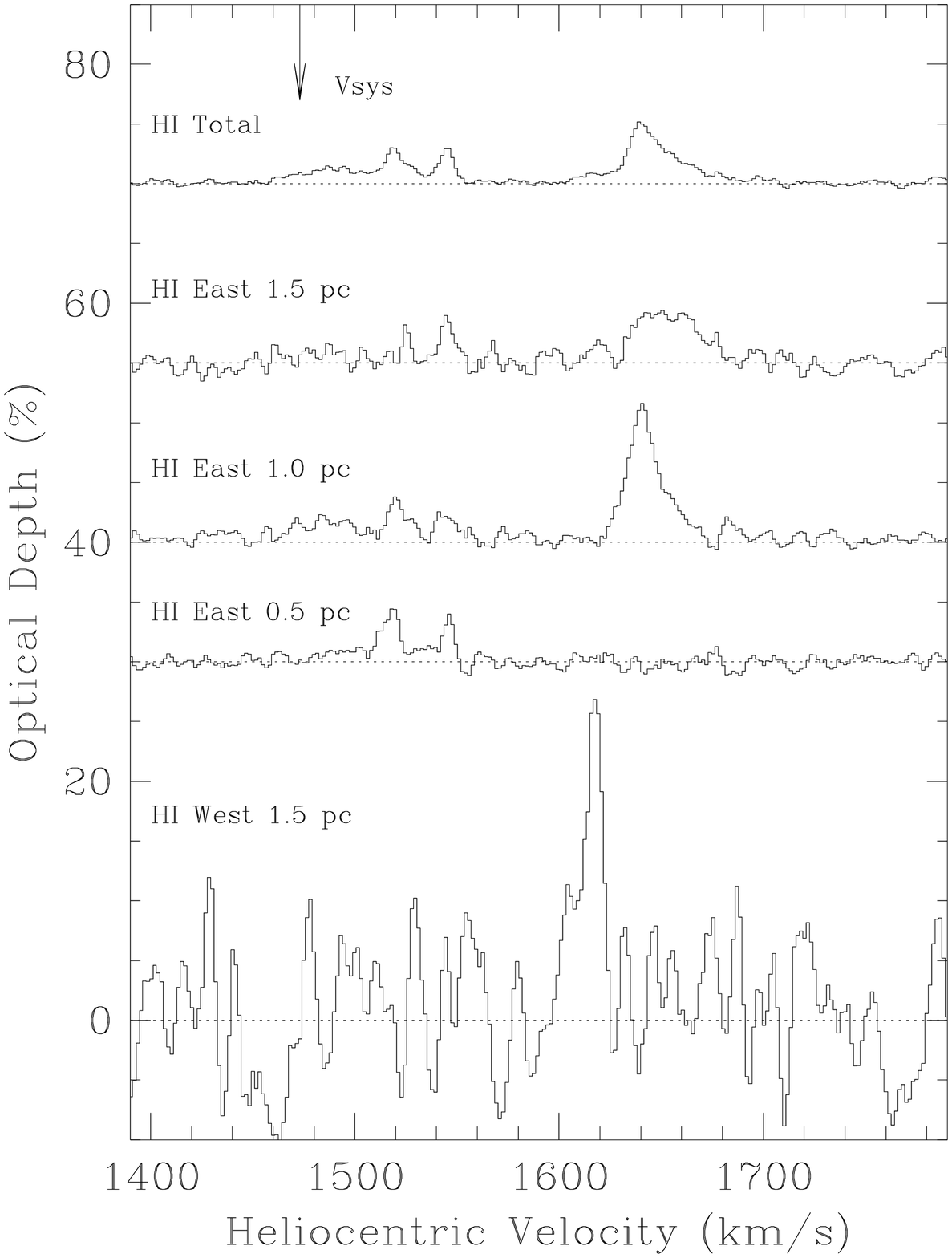}
\vspace{-15pt}
\end{center}
\caption{H\,{\sc i} optical depths as function of velocity, at various
locations along the jets of \Index{NGC\,1052}, as observed with VLBI in July
1998.  Offsets between the spectra are used for clarity; dotted lines
show the zero levels.
\textsl{Taken from Vermeulen et al.\, (2003, A\&A, 401, 113).}
}
\end{figure}

\paragraph{4.} WSRT observations show, from 1667 and 1665\,MHz OH main lines seen
in absorption, that molecular gas exists along the full velocity span
found also in atomic gas, roughly $-30$ to 200\,km\,s$^{-1}$. The peak
absorption depth, about 0.4\%, occurs in the ``high velocity'' system,
and indicates a column depth of order $10^{14}$\,cm$^{-2}$. The ratio
between the OH 1667 and 1665\,MHz lines is about 2 in the ``high
velocity'' system, but closer to 1 at lower velocities. In the ``high
velocity'' system, the 1612 and 1720\,MHz OH satellite lines have also
been detected, both with a peak strength near 0.25\%, but with
conjugate profiles, in absorption and emission, respectively. The
profiles of the OH main lines and the integrated H\,{\sc i} line are
remarkably similar in the ``high velocity'' system, which suggests
co-location of the atomic and molecular gas at these velocities. But
H$_2$O masers, which cover the same velocity range, were found very
close to the nucleus, at 0.1--0.2\,pc along the receding jet (Claussen
et al.\ 1998) We have no satisfactory model to explain this, in light
of the central hole in H\,{\sc i}, the OH and H\,{\sc i} profile
similarity, the velocity gradient across the core (suggestive of
rotation), and the substantial overall velocity offset compared to
systemic (suggestive of infall).

\printindex
\end{document}